# Systematic Reconstruction of Disease Networks from Longitudinal Blood Data for Causal Discovery and Intervention Analysis

**David Patrick Duys Montealegre[1], Alexander Fulton[2,3], Mahta Haghighat Ghahfarokhi[1,6], Abicumaran Uthamacumaran[4], Hector Zenil[1,4,5,6,7]***

[1]Algorithmic Dynamics Lab, King's Institute for Artificial Intelligence, King's College London, UK
[2] Cambridge University NHS Foundation Trust, Cambridge, UK
[3] Precision Breast Cancer Institute and Department of Oncology, University of Cambridge, Cambridge, UK
[4] Cancer Research Interest Group, The Francis Crick Institute, London, UK
[5] Oxford Immune Algorithmics, Oxford University Innovation & London Institute for Healthcare Engineering, UK
[6] Departments of Biomedical Computing and Digital Twins, School of Biomedical Engineering and Imaging Sciences, King's Faculty of Life Sciences and Medicine, King's College London, UK
[7] EPSRC Future Blood Testing Network+ and King's Health Partners AHSC (NHS–KCL), UK

## Abstract

We explore the hyperparameters and introduce a methodological framework to convert disease patterns from time series data of blood test results into correlation graphs for causal hypothesis exploration. The networks represent hypotheses that can then be validated or rejected both for causal discovery and causal analysis (under intervention). We synthetically recreated a repository of 105 typical disease longitudinal patterns extracted from medical guidance and research literature of common blood markers to build a systematic pipeline to translate multidimensional clinical data into intervenable disease networks for causal discovery and causal analysis. This study demonstrates that knowledge graphical models reconstructed from longitudinal data can transform routine medical data into clinically interpretable structures. By integrating multiple thresholding strategies and causal graph design, the framework has the purpose to move beyond statistical correlation toward clinically and testable inference networks. These results highlight a practical pathway for more transparent, explainable, and scalable tools in clinical decision support for AI training, precision healthcare and predictive medicine, offering interpretable, clinically actionable outputs that support safer use of AI in differential diagnosis.

* Corresponding author: hector.zenil@kcl.ac.uk

## 1. Introduction

Causal inference (CI) in medicine identifies cause–effect relationships to improve diagnosis, evaluation, and prevention by moving beyond correlations in heterogeneous data. Traditional statistical methods often fail to interpret complex markers or handle confounding, whereas CI allows more accurate, reasoning-based approaches (Sanchez et al., 2022). Current clinical decision support (CDS) tools "learn only associations between variables … without distinguishing causal relationships and (spurious) correlations," limiting actionable insights (Sanchez et al., 2022). Pearl (2009) argued causation enriches statistics by uncovering mechanisms beyond correlations. Integrating AI with CI thus enhances CDS by synthesizing data, automating diagnostics, and improving predictive tasks across images, notes, and tests (Sanchez et al., 2022). Structural Causal Models (SCMs), represented as directed acyclic graphs (DAGs), estimate causal effects while addressing spurious associations, improving prediction (Pearl, 2009; Sanchez et al., 2022). CI discovery algorithms further enable automated learning from observational data, strengthening inference in limited or noisy contexts (Kalainathan et al., 2019).

Knowledge graphs represent biomedical entities as nodes and relationships as edges, enabling visualization of complex disease mechanisms (Renaux, 2023). They enhance structuring of clinical knowledge using ontologies such as UMLS (Bodenreider et al., 2004) and integrate diverse datasets for hypothesis generation and decision support (Zhang et al., 2020). Despite their utility, graphs remain underused in real-time diagnosis due to static nature and limited integration with machine learning workflows (Zhang et al., 2020). Combining knowledge graphs with CI and generative AI could enable dynamic, interpretable, and context-aware reasoning in clinical settings (Peng et al., 2023).

RAG augments large language models by retrieving external knowledge, producing factually grounded outputs and reducing hallucinations (Lewis et al., 2020). In medicine, this supports factuality, comprehension, and reasoning benchmarks, though current models remain inferior to clinicians (Singhal et al., 2023). Refinements such as Med-PaLM and BioMED-RAG integrate PubMed and case repositories, yielding cited, accurate responses (Singhal et al., 2023). Coupling RAG with causal knowledge graphs further enables counterfactual reasoning for differential diagnosis (Singhal et al., 2023). Unlike parametric memory models, RAG retrieves from external databases and graphs (Karpas et al., 2022; Izacard et al., 2022), aligning outputs with literature and patient safety requirements.

AI integration can optimize decision support and diagnosis, but current models prioritize predictive accuracy over interpretability. Black-box deep learning constrains causal reasoning (Rajkomar et al., 2019; Shiwlani et al., 2024), while pattern-identification without mechanistic understanding yields unstable, non-generalizable outputs (Obermeyer & Emanuel, 2016; Caruana et al., 2015; Topol, 2019). This limits patient-specific inference, especially where distinct conditions share similar Full Blood Count or FBC (also known as Complete Blood Count or CBC) profiles (Miotto et al., 2016). Since diagnosis is inherently a causal task, current AI's lack of counterfactual reasoning hinders its clinical utility (Pearl, 2009; Zhao et al., 2020).

Heterogeneity in medical data (structured tests, notes, variable standards) further complicates integration, requiring adaptive preprocessing and harmonization (Miotto et al., 2016; Torab-Miandoab et al., 2023). Biomedical knowledge graphs remain static and poorly aligned to evolving patient data, limiting real-time use (Schriml et al., 2019; Hogan et al., 2021). RAG systems, while promising, rely on general embeddings not tailored to complex ontologies, limiting interpretability and causal reasoning (Lewis et al., 2020; Alsentzer et al., 2019; Pearl & Mackenzie, 2018).

Fragmentation compounds these issues: CI, knowledge graphs, and AI frameworks evolve in isolation, lacking interoperability and integration into real-world CDS (Weissler et al., 2021; Shortliffe & Sepulveda, 2018; Topol, 2019). This absence of synthesis increases complexity and clinician burden, undermining trust and adoption.

Generative AI models such as GatorTron (Yang et al., 2022), BioGPT (Luo et al., 2022), and Med-PaLM (Singhal et al., 2023; Lee et al., 2020) demonstrate strong performance in NLP and Q&A tasks but risk hallucinations and lack causal reasoning (Lewis et al., 2020; Shortliffe & Sepulveda, 2018). Causal machine learning, including forests and Bayesian networks, enables treatment effect estimation and patient stratification (Sanchez et al., 2022), while neural discovery models such as CGNN show promise in nonlinear, multimodal settings (Brouillard et al., 2020). Clinical knowledge graphs (Bodenreider et al., 2004; Schriml et al., 2019) provide structured biomedical relationships but remain underutilized for personalized reasoning, though integration with GNNs and reasoning engines shows potential.

## 2. Methodology

Here we develop an integrated pipeline towards causal inference and knowledge graphs from longitudinal clinical data and blood test time series. The pipeline comprises of (i) data preprocessing, (ii) knowledge graph construction, and (iii) correlation- and causal-style graph generation. Full blood count (FBC) also known as Complete Blood Count (CBC) data values manually curated from medical guidance served as primary input for this disease-network reconstruction framework.

### 2.1 FBC Data Collection, Preprocessing, and Refinement

We assembled 105 synthetic hematological datasets with FBC analytes of simulated patients with 105 different diseases over time. As illustrated in **Table 1** (Microcytic Anaemia), analytes were measured longitudinally (Days 0, 7, 14, 21, 28), enabling trajectory mapping and inter-analyte relationship analysis. Profiles were ingested from a multi-sheet workbook and standardized for graph-based modeling. Sheet names corresponding to individual clinical profiles were programmatically enumerated and iterated to extract FBC data. Analytes were assigned distinct markers/shapes to improve node-level interpretability in graphs. Non-numeric headers, free text, and metadata were excluded to retain analyte-level numeric matrices. Rows with all-zero or all-NaN values were removed as a filtering step; sparse within-row gaps were linearly interpolated when ≥2 timepoints were present, preserving analytes with minor missingness *(see Data and Code Availability section)*. To ensure clinical relevance, we benchmarked temporal variability against known within-analyte normal biological variation ($BV_{intra}$). For each analyte, the observed longitudinal standard deviation ($SD_{profile}$) was required to meet a minimum variability threshold ($SD_{profile} \geq 0.05 \times BV_{intra}$), ensuring only analytes with meaningful fluctuations were analyzed. Additionally, the total White Blood Cell (WBC) count was excluded from correlation analysis to prevent trivial redundancy with its constituent subsets (neutrophils, lymphocytes, etc.).

**2.2 Correlation Graph Construction and Threshold Optimization**

Pairwise correlations were computed using both Pearson (r) and Spearman (ρ) coefficients to capture linear and rank-based monotonic relationships within each profile to quantify longitudinal analyte relationships *(see Data and Code Availability section)*. Pearson was used for primary structural modeling, and Spearman coefficients provided robust descriptive labels (Strong: $|\rho| \geq 0.90$; Moderate: $0.70 \leq |\rho| < 0.90$) and sensitivity analysis.

To focus on salient structure and reduce noise, we evaluated three complementary thresholding strategies and selected the most conservative cut-off per profile:

- **Natural threshold:** inflection-based cutoff using Gaussian mixture modeling of $|r|$, capturing structural stability (Zhang et al., 2021; Toubiana et al., 2021).
- **Elbow method**: identifies the maximum curvature in the sorted $|r|$ distribution to define the cutoff (Atunes et al., 2025).
- **Ratio method:** retains the top 30% strongest $|r|$ values, a conservative hyperparameter applied where curvature is weak.

The chosen threshold defined an undirected, weighted correlation graph: nodes are analytes; edges connect pairs with $|r| \geq$ threshold; edge widths scale with $|r|$ *(see Data and Code Availability section)*. Graphs were exported with node coordinates, edge weights, and threshold metadata for each profile.

To statistically assess the discrimination achieved by thresholding and any disagreements between the two methods used for correlation calculation, correlation edges were divided into retained ($|r| \geq$ natural threshold, $|\rho| \geq$ natural threshold) and discarded ($|r| <$ threshold, $|\rho| <$ threshold) for both Pearson and Spearman. The distributions of absolute correlations were compared using the Mann–Whitney U test and Welch's unequal-variance t-test, with effect size quantified by Cohen's d. This approach evaluates whether thresholding systematically preserves stronger associations while excluding weaker, potentially spurious correlations.

To further validate these associations, we generated empirical null distributions via permutation testing. For each profile, time points were randomly permuted 5000 times per analyte to break temporal structure while preserving marginal distributions. We computed the 95th and 99th percentiles of the maximum absolute correlations from these permuted datasets. Valid associations were required to exceed this profile-specific empirical 95th percentile threshold, ensuring they were statistically distinct from random noise.

Robust analyte–analyte associations were defined using a conservative consensus approach. To be retained in the final graph, an association was required to simultaneously satisfy the Natural, Elbow, and Ratio thresholds, as well as exceed the empirical 95th percentile permutation threshold. This approach ensured that only structurally salient and statistically significant edges were preserved.

**2.3 Causal-Style Graph Construction**
To approximate latent mechanistic structure without imposing analyte-to-analyte directionality, we generated causal-style graphs from the thresholded correlations. For each strong pair, a shared latent "cause" node was introduced with directed edges to the two analytes, representing a common-cause hypothesis consistent with causal diagram conventions (Pearl, 2009). This yields a directed acyclic representation centered on latent drivers while avoiding unsupported direct causal claims between analytes *(see Data and Code Availability section)*. The resulting graphs highlight candidate latent events and prioritize analyte interactions for subsequent structural causal modeling and counterfactual testing.

| Disease Profile | Microcytic Anaemia | | | | |
|---|---|---|---|---|---|
| Scenario | IDA secondary to dietary deficiency | | | | |
| Symptoms* | Fatigue, pallor, dyspnoea, pica. Other symptoms may be present depending on cause such as bleeding or malignancy related symptoms. | | | | |
| *Additional symptoms possible, this list is to highlight some of the common presenting features. | | | | | |
| Associated Guidelines: BMJ Best Practice | | | | | |
| Comments | Routine blood test with GP | D7 - IV iron replacement | D14 - second iron infusion | D21 | D28 |
| FBC | | | | | |
| WBC (1000 cells/uL) | 6.5 | 6.1 | 6.6 | 6.1 | 5.7 |
| Lymphocytes (1000 cells/uL) | 2 | 1.9 | 2.1 | 2 | 1.8 |
| Monocytes (1000 cells/uL) | 0.2 | 0.1 | 0.1 | 0.2 | 0.1 |
| Segmented neutrophils (1000 cells/uL) | 4.2 | 4.1 | 4.4 | 3.9 | 3.8 |
| Eosinophils (1000 cells/uL) | 0.1 | 0 | 0 | 0 | 0 |
| Basophils (1000 cells/uL) | 0 | 0 | 0 | 0 | 0 |
| RBC (million cells/uL) | 4 | 3.9 | 4.1 | 4.2 | 4.4 |
| Haemoglobin (g/dL) | 11.8 | 11.7 | 12 | 12.3 | 12.7 |
| MCV (fL) | 76 | 75 | 78 | 79 | 81 |
| Platelet count (1000 cells/uL) | 236 | 247 | 272 | 266 | 300 |
| MPV (fL) | 7.8 | 7.6 | 7.5 | 7.8 | 7.6 |
| Renal Function | | | | | |
| Na (mmol/L) | 139 | 138 | 135 | 137 | 140 |
| K (mmol/L) | 3.7 | 3.9 | 3.7 | 4 | 3.8 |
| Urea (mmol/L) | 5.1 | 5 | 4.8 | 5 | 4.8 |
| Creat (mmol/L) | 70 | 73 | 69 | 71 | 75 |
| CRP (mg/L) | <4 | <4 | <4 | <4 | <4 |

***Table 1: Example of an Haematological FBC Clinical Profile for Microcytic anaemi..***

## 3. Results

A total of 105 clinical profiles were processed using the model described in Section 2. Each profile included FBC analyte values across 9 variables: WBC (1000 cells/uL), Lymphocytes (1000 cells/uL), Monocytes (1000 cells/uL), Segmented Neutrophils (1000 cells/uL), RBC (million cells/uL), Haemoglobin (g/dL), MCV (fL), Platelet Count (1000 cells/uL), MPV (fL). For each profile, correlation matrices were computed, thresholded using three methods (natural, elbow, and ratio), and used to generate both correlation graphs where the passed correlations are dynamically adjusted with respect to the thresholding value, and causal graphs with intermediate red causal nodes connecting strongly associated analytes.

Graphs were generated for every profile and saved in CSV format, including nodes, edge weights, and threshold metadata. As seen below, these are the computed correlation values for each analyte pair, assorted in strongest to weakest. Moreover, the table below also records all 3 thresholding calculations to demonstrate the yielded estimations and which of these are the most conservative.

| **Analyte Pair** | **Correlation** |
|---|---|
| RBC (million cells/uL) vs Haemoglobin (g/dL) | 0.99 |
| RBC (million cells/uL) vs MCV (fL) | 0.99 |
| Haemoglobin (g/dL) vs MCV (fL) | 0.97 |

| Monocytes (1000 cells/uL) vs MPV (fL) | 0.952 |
|---|---|
| MCV (fL) vs Platelet count (1000 cells/uL) | 0.921 |
| Haemoglobin (g/dL) vs Platelet count (1000 cells/uL) | 0.907 |
| RBC (million cells/uL) vs Platelet count (1000 cells/uL) | 0.904 |
| Lymphocytes (1000 cells/uL) vs Segmented neutrophils (1000 cells/uL) | 0.789 |
| Monocytes (1000 cells/uL) vs Eosinophils (1000 cells/uL) | 0.612 |
| Eosinophils (1000 cells/uL) vs MPV (fL) | 0.583 |
| Lymphocytes (1000 cells/uL) vs Monocytes (1000 cells/uL) | 0.32 |
| Segmented neutrophils (1000 cells/uL) vs Eosinophils (1000 cells/uL) | 0.28 |
| Lymphocytes (1000 cells/uL) vs Eosinophils (1000 cells/uL) | 0.196 |
| Lymphocytes (1000 cells/uL) vs MPV (fL) | 0.032 |
| Haemoglobin (g/dL) vs MPV (fL) | -0.045 |
| RBC (million cells/uL) vs MPV (fL) | -0.058 |
| Monocytes (1000 cells/uL) vs RBC (million cells/uL) | -0.094 |
| MCV (fL) vs MPV (fL) | -0.1 |
| Monocytes (1000 cells/uL) vs Haemoglobin (g/dL) | -0.112 |
| Monocytes (1000 cells/uL) vs Segmented neutrophils (1000 cells/uL) | -0.114 |
| Monocytes (1000 cells/uL) vs MCV (fL) | -0.114 |
| Lymphocytes (1000 cells/uL) vs MCV (fL) | -0.312 |
| Segmented neutrophils (1000 cells/uL) vs MPV (fL) | -0.343 |
| Eosinophils (1000 cells/uL) vs RBC (million cells/uL) | -0.348 |
| Lymphocytes (1000 cells/uL) vs RBC (million cells/uL) | -0.41 |
| Eosinophils (1000 cells/uL) vs Haemoglobin (g/dL) | -0.412 |
| Lymphocytes (1000 cells/uL) vs Platelet count (1000 cells/uL) | -0.413 |
| Eosinophils (1000 cells/uL) vs MCV (fL) | -0.421 |
| Platelet count (1000 cells/uL) vs MPV (fL) | -0.457 |
| Lymphocytes (1000 cells/uL) vs Haemoglobin (g/dL) | -0.485 |
| Monocytes (1000 cells/uL) vs Platelet count (1000 cells/uL) | -0.488 |
| Segmented neutrophils (1000 cells/uL) vs Platelet count (1000 cells/uL) | -0.491 |
| Segmented neutrophils (1000 cells/uL) vs MCV (fL) | -0.578 |
| Eosinophils (1000 cells/uL) vs Platelet count (1000 cells/uL) | -0.638 |
| Segmented neutrophils (1000 cells/uL) vs RBC (million cells/uL) | -0.642 |
| Segmented neutrophils (1000 cells/uL) vs Hemoglobin (g/dL) | -0.721 |
| **Description** | **Value** |
| Natural Threshold | 0.480 |

| Elbow Threshold | 0.491 |
|---|---|
| Ratio Threshold | 0.782 |
| Permutation Threshold 95th Percentile | 0.854 |
| Threshold Used (Permutation Threshold 95th Percentile) | 0.854 |

***Table 2: Example Estimation of Correlation Values and Threshold Results for the Microcytic Anaemia Profile***

Employing these computed values, the model constructed correlation graphs according to the selected threshold and adjusted the correlation values as training progressed on the profile. In this example, thresholding retained 10 edges ($|r| \geq 0.797$) and discarded 26. The mean absolute correlation of retained edges was substantially higher than discarded edges (0.946 vs. 0.377). Shapiro–Wilk testing indicated non-normality in the retained group ($p < 0.05$) and borderline normality in the discarded group ($p = 0.073$). Non-parametric analysis confirmed significantly stronger correlations among retained edges (Mann–Whitney $U = 260$, $p < 0.001$), supported by Welch's t-test ($t = 15.86$, $p < 0.001$) and a very large effect size (Cohen's $d = 4.83$). These results demonstrate that thresholding robustly distinguishes high-strength associations from weaker or spurious correlations, reinforcing its utility for generating clinically interpretable graphical structures.

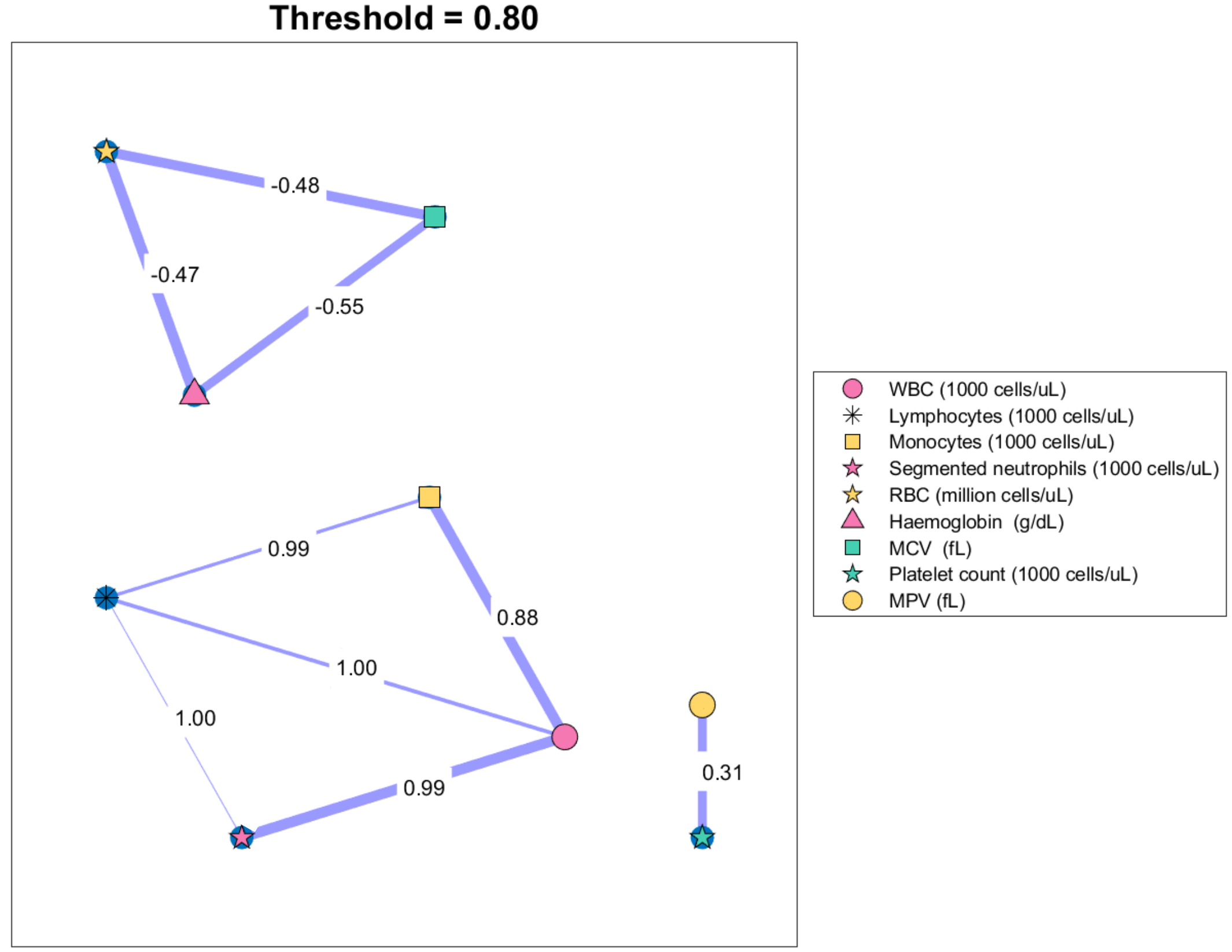

*Figure 3: Correlation Example Graph with Thresholding for the Profile 2–Microcytic Anaemia from 105 Total Disease Profiles*

The correlation graph, however, still retains associations that are not clinically significant or that align with medical literature. Employing the knowledge graph on this specific clinical profile (Microcytic anaemia), the causal graph trains this grounded and medical information to produce a more accurate causal structure more closely aligning to the accepted literature. As such, as seen below, the causal graph severs edges between analytes it considers as spurious associations or negligible relationships that are not medically grounded. Moreover, it recalculates the correlations over a more conservative dynamic range and stricter thresholding estimation.

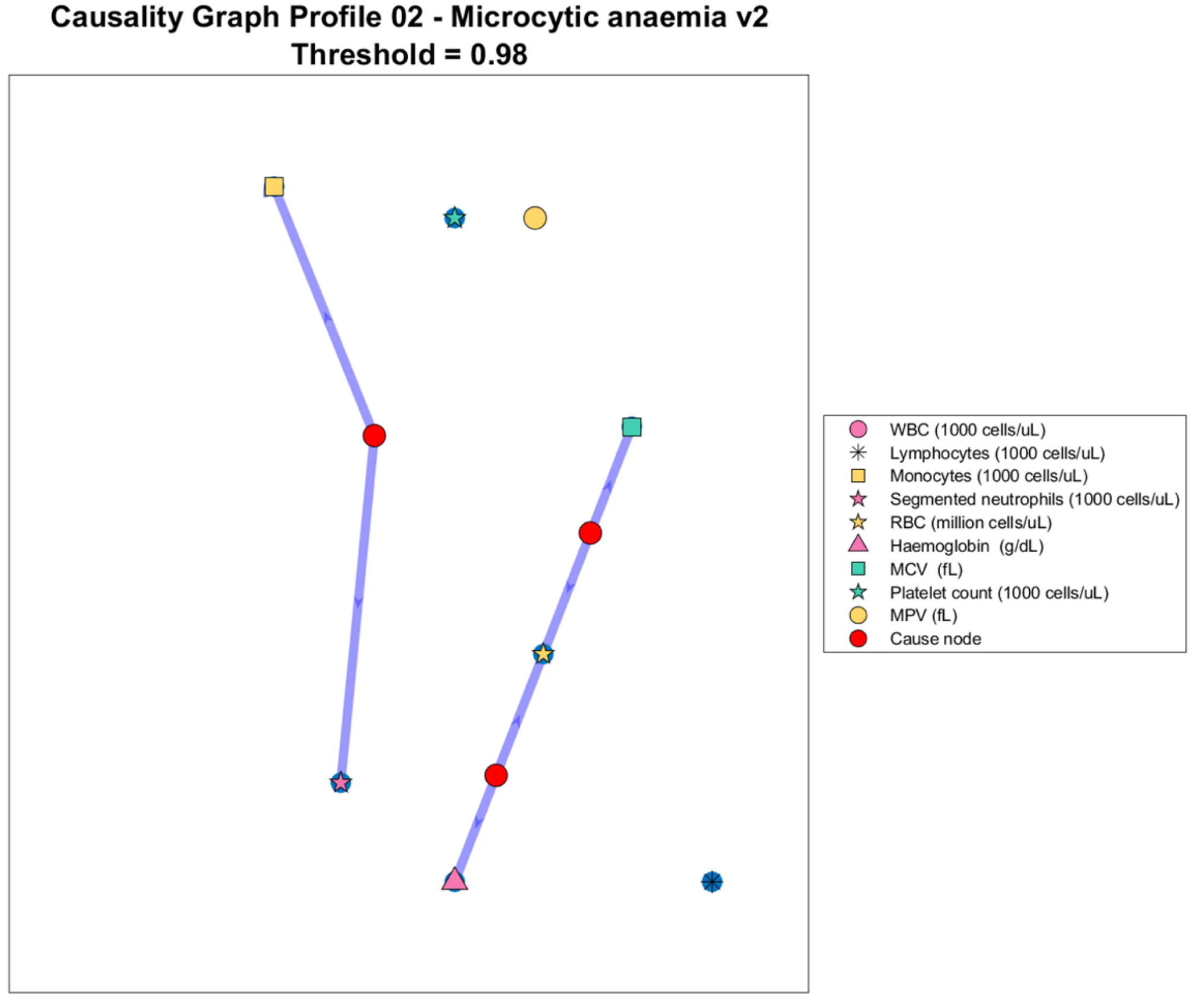

*Figure 4: Causal Graph with Refined Thresholding and Cause Nodes for Profile 2–Microcytic Anaemia Case Showing Multiple Disconnected Causal Hypotheses*

Additionally, the causal graph includes cause nodes, evidenced by the red nodes placed between analytes that exhibit strong correlations, with a causal directionality emanating from these nodes to the analytes. This enables the model to visualize the causal nature between analytes without assigning specific directionality, enabling preliminary use of the model in clinical settings.

## 4. Discussion

The construction of both correlation and causal graphs iterated for all 105 distinct clinical profiles employed throughout this project yield insightful information into the causal structures and analyte relationships for differing disease profiles. Moreover, the employment of different thresholding and modelling techniques output different results that highlight the complexity in this project at producing results affecting clinical plausibility but also interpretability. This provision of dual-layered representation of FBC analyte relationship enables the deepened comprehension between the correlation and mechanisms of analytes while also understanding the medical significance each has when cross-referenced with grounded literature.

### 4.1 Comparison of Graph Dynamics and Clinical Coherence

The initial correlation graphs captured statistical co-expression patterns within each clinical profile, allowing us to establish significant associations and train the model on these filtered relationships. The refined causal graph introduced a layer of interpretability grounded in clinically plausible reasoning and ensured the retained associations align with factual literature, optimizing interpretability for the clinical workflow. By contrasting both modelling techniques while employing varying thresholding approaches (natural, elbow, and top-30% ratio), the study highlights trade-offs between completeness and clarity, statistical strength and medical correctness, and ultimately, correlation and causation.

A key observation is that although correlation graphs demonstrate analyte pairs with high-strength edges, many do not align with established literature and can be attributed to spurious associations, noise, limited training, or insufficient data. This limitation of purely statistical models, which tend to overfit and emphasize approximation over factualness, contrasts with causal graphs enhanced with domain-informed constraints and conservative thresholds. In the Microcytic Anaemia profile (Figure 4), edges in the causal graph better mirrored known associations, such as RBC–haemoglobin, while weaker links were omitted.

Correlation graphs, despite conservative thresholds, retained many relationships, producing densely connected structures. For example, the Microcytic Anaemia correlation graph (Figure 3) yielded nearly twice as many edges as its causal counterpart. Across 105 profiles, causal graphs reduced connectivity by ~27.4% relative to correlation graphs, preserving only the strongest associations. In the Iron Deficiency Anaemia (IDA) profile, the correlation graph retained links between clinically unrelated analytes, while the causal graph, trained on knowledge graphs and benchmark literature, filtered these out and preserved established patterns such as RBC–haemoglobin (0.86) and MCH–MCV (0.81), consistent with IDA pathophysiology.

This reduction in redundancy was consistent: correlation graphs retained 27.4% more connections on average, many lacking factual support. Profiles such as Leukemoid Reaction and Chronic Inflammation showed the most pronounced reductions (~40%) when moving from correlation to causal representation, reflecting their greater prevalence of co-fluctuating analytes and spurious associations likely driven by systemic inflammatory perturbations.

In the causal graphs, each strongly associated analyte pair above the threshold is indirectly connected through a common cause node (Figure 4). This design frames high-correlation relationships as effects of latent factors or non-directional mechanisms not yet clinically understood, replacing direct mechanistic links with proxy structures. Such representation supports exploration of alternative clinical explanations and distinguishes between mere statistical associations, as in most machine learning tools, and clinically plausible interactions. The model therefore eliminates spurious or clinically insignificant connections despite statistical strength, yielding medically relevant insights.

Across 105 clinical profiles, causal graphs preserved strong, literature-supported interactions while systematically filtering weaker associations present in correlation-only models. Training on knowledge graphs specific to each condition enabled cross-referencing between significant correlations and accepted biomedical relationships, grounding results in published evidence. Following Pearl's (2009) framework, this structure highlights interventions and mediating variables rather than simple covariation. For example, in the Megaloblastic Anaemia profile, strong associations between MCV and haemoglobin, medically explained by B12 deficiency, appear in the causal graph via a shared intermediary node, explicitly representing the biological driver. This design enhances interpretability, hypothesis generation, and positions the model for integration with formal structural causal modeling tools that combine statistical inference and domain knowledge.

### 4.2 Comparison of Thresholding Techniques

Thresholding strategies fundamentally determined graph topology, revealing a significant divergence between distribution-based heuristics and rigorous statistical validation. The Natural Threshold and Elbow Method proved structurally permissive, yielding mean edge counts of 10.15 (±3.55) and 9.54 (±4.95) per profile, respectively. These methods effectively capture broad structural dependencies but risk retaining lower-magnitude associations driven by stochastic noise.

In contrast, the Permutation Threshold (95th percentile)—which controls the Family-Wise Error Rate (FWER) by benchmarking against the distribution of maxima from randomized data—imposed a rigorous constraint, reducing mean connectivity to 3.94 (±3.21) edges per profile.

This discrepancy highlights a critical finding: distribution-based heuristics (Natural/Elbow) overestimated clinically relevant connectivity by a factor of ≈2.5 compared to the statistical null model. For example, in the Microcytic Anaemia profile, the Natural threshold (0.480) would have retained weak associations indistinguishable from random fluctuations. However, the Permutation threshold (0.854) successfully filtered these, retaining only the strongest signal.

Notably, the Ratio Method (mean 3.69±1.82 edges) closely approximated the statistical ceiling established by the Permutation test. This validates the utility of our conservative consensus approach. By requiring associations to simultaneously satisfy heuristic criteria (Natural/Elbow) and exceed the strict Permutation threshold, the framework effectively suppressed the ≈60% of edges captured by weaker methods that lacked statistical significance. This prioritization of specificity over sensitivity is a requisite safeguard for causal discovery in clinical contexts, ensuring that generated networks reflect robust biological signals rather than spurious correlations.

### 4.3 Limitations

Reliance on Pearson correlation restricts detection of non-linear, monotonic, or conditional relationships essential for longitudinal mapping. Incorporating Spearman, mutual information, or kernel-based metrics would expand non-linear capabilities and improve generalizability.

Training used synthetic profiles that, while controlled and tailored to known FBC patterns, lack the noise, outliers, and confounders present in real hospital data. This limits ecological validity and requires future training and validation with empirical EHR and laboratory datasets.

Natural and elbow methods assume unimodal correlation decay, which does not generalize across all pathologies. In multimodal profiles, the elbow method sometimes produced thresholds that overemphasized sparsity at the expense of clinical relevance. For conditions such as inflammatory profiles where platelet dynamics dominate, higher-order or conditional models may better capture underlying dependencies.

The framework treats all correlations as symmetric and undirected between analytes and common cause nodes, limiting causal inference. While cause nodes aid interpretability, they do not provide directionality or conditional independence testing. Future iterations should incorporate robust causal discovery to assign direction, model latent confounding, and quantify influence on outcomes.

The construction of causal graphs for CDS, while promising, must be approached with diligence given ethical risks. Overinterpreting causal links, omitting associations, and the limited real-world testing of such models raise concerns about misuse and potential harm in differential diagnosis. Misdiagnosis and bias may inhibit acceptance, so rigorous validation and monitoring are essential to ensure safety and to confirm that implementation yields net benefit rather than workflow disruption.

## 5. Conclusion and Future Work

This study addresses a critical gap in CDS: the absence of interpretable, dynamic, context-aware, and causal structures for haematological data. To meet this need, we developed a framework for generating causal-style graphical models from FBC disease profiles, emphasizing explainability, transparency, and clinical grounding. By employing multiple thresholding methods tailored to specific profiles, along with correlational and causal visualization techniques, the framework optimizes interpretability and preserves medically significant analyte relationships while minimizing spurious complexity, statistical noise, and unsupported associations.

Results demonstrate that the causal graph framework consistently produced clinically grounded structures across a sample of 105 pathological FBC profiles, aligning closely with established relationships through integration with knowledge graphs and improving upon purely correlational approaches. Key limitations include reliance on synthetic data curated by a human clinical domain expert, the assumption of linear correlations, restricted thresholding in multimodal contexts, and the treatment of correlations as symmetric and non-directional.

Future work should extend this framework to dynamic longitudinal settings, enabling modelling of evolving haematological trajectories, expanding beyond FBC and using more powerful causal discovery and causal analysis tools such as Algorithmic Information Dynamics (Zenil 2019; Zenil et al 2023). Such advances would bridge current statistical methods with causal AI capabilities even beyond Judea Pearl's statistical causal inference, enhancing diagnostic utility and CDS while minimizing workflow disruption by incrementing mechanistic automation from first principles without human close supervision.